\documentclass[twoside
]{elsart3p}
\usepackage{color} 
\usepackage{times} 
\usepackage{mathptm} 
\usepackage{graphicx} 
\journal{Astroparticle Physics} 
\def\pp#1{pp\if#1\ { }\else{#1}\fi} 
\def\be#1{\ensuremath{^{#1}{\rm Be}}} 
\def\Be#1{\ensuremath{^{#1}{\rm Be}}} 
\def\he#1{\ensuremath{^{#1}{\rm He}}} 
\def\B#1{\ensuremath{^{#1}{\rm B}}} 
 
\def\au{\ensuremath{(\rm A.U.)}} 
\def\phipp{\ensuremath{\Phi_{\rm pp}}} 
\def\phibe{\ensuremath{\Phi_{\rm Be}}} 
\def\sun{{\ensuremath{\odot}}} 
 
\def\electron{{\rm e}} 
\long\def\symbolfootnotemark[#1]{\begingroup%
\def\thefootnote{\fnsymbol{footnote}}\footnotemark[#1]\endgroup} 
\long\def\symbolfootnotetext[#1]#2{\begingroup%
\def\thefootnote{\fnsymbol{footnote}}\footnotetext[#1]{#2}\endgroup}

\begin{document} 
\begin{frontmatter} 

\title{Potential for Precision Measurement of Solar Neutrino Luminosity by 
HERON}

\author{Y.H.  Huang}, 
\author{R.E.  Lanou},\symbolfootnotemark[1] 
\ead{lanou@hep.brown.edu} 
\corauth{Corresponding author.} 
\author{H.J.  Maris}, 
\author{G.M.  Seidel}, 
\author{B.  Sethumadhavan},\symbolfootnotemark[2] 
and \author{W.  Yao}\symbolfootnotemark[3]
\address{Department of Physics, Brown University, Providence, 
RI 02912, USA}

\begin{abstract} 
Results are presented for a simulation carried out to test the precision
with which a detector design (HERON) based on a superfluid helium target
material should be able to measure the solar \pp\ and \be7 fluxes.  It is
found that precisions of $\pm 1.68\%$ and $\pm 2.97\%$ for \pp\ and \be7
fluxes, respectively, should be achievable in a 5-year data sample.  The
physics motivation to aim for these precisions is outlined as are the
detector design, the methods used in the simulation and sensitivity to solar
orbit eccentricity.
\end{abstract} 
\end{frontmatter} 

\symbolfootnotetext[2]{Now at: Intel Corporation.} 
\symbolfootnotetext[3]{Now at: Magnet Laboratory, Mass.  Inst.  of Tech.} 

\section{Introduction} 
This paper presents an analysis of the projected capability of a detector
design, HERON \cite{heron}, based on a target material of superfluid helium
to make a precise measurement of both the \pp\ and \be7 solar neutrino
fluxes (\phipp{} and \phibe, resp.) in a single, real-time experiment.  The
detection reaction used would be the elastic scattering of neutrinos by
electrons (ES).  In addition to the novel use of helium, the detector also
includes the novel application of a coded aperture \cite{dicke,ura}
technique for accurate measurement of the location and recoil energy of each
elastic scattering event and to aid in background discrimination.  According
to models of the Sun \cite{bethe}, the neutrinos from the \pp-I and \pp-II
branches of the fusion chain (known as the \pp\ and \be7 neutrinos,
respectively) are, when taken together, $> 98\%$ of the neutrino flux and
are associated with the reactions producing a similar fraction of the solar
energy.  At the present writing there have been no real-time experiments to
measure the flux and spectra of \phipp{} but recently the Borexino
collaboration has made the first real-time spectral detection of \phibe
($861\,{\rm keV}$) \cite{borexino}.  As we explain in Sec.~\ref{sec:goal}
there are several important physics issues which can be addressed if a
detector can be constructed to measure both these fluxes and spectra with
sufficient precision.  For the \pp\ neutrinos ($<420\,$keV) there does not
yet exist any detector with demonstrated feasibility to measure either their
flux or spectra; however, there are a number of efforts \cite{heron,exps}
which aim to do so.

Section~2 of this paper discusses the physics goals motivating the HERON
detector.  Sec.~3 briefly presents the requirements for and description of
the HERON detector design.  Sec.~4 and 5 provide the details of the HERON
capability analysis.  In an Appendix we discuss an application to measuring
the solar orbit eccentricity.

\section{Physics Goals\label{sec:goal}} 
A principal goal of HERON would be to make an accurate measurement of the
luminosity of the Sun using precise measurements of active neutrino fluxes.
An experiment capable of measuring both \phipp{} and \phibe{} sufficiently
well for an accurate luminosity measurement can also address several other
interesting topics.  These include testing for the relative rates of the
\he3(\he4, 2p)\he4 and \he3(\he4, $\gamma$)\be7 reactions which terminate 
the \pp-I and \pp-II branches in the Sun and also for testing the MSW effect
(after Mikheyev, Smirnov, Wolfenstein \cite{msw}) in the LMA (large mixing
angle) solution to the ``solar neutrino problem''.  Additionally, if new
measurements of \phipp{} and \phibe{} are successful at the few-percent
level then, via new, luminosity-constrained global fits to all neutrino
data, some modest improvement can be made in the knowledge of
$\tan^2\theta_{12}$, $\sin^2\theta_{13}$ and limits on sterile neutrinos.
Lastly, since a real-time low-energy solar neutrino experiment opens a new
window in neutrino physics, the possibility of surprises in the physics
should not be discounted.

\subsection{Solar Luminosity.  Why measure the solar luminosity by 
neutrinos?} The radiant photon energy reaching the Earth from the Sun (the
irradiance, $I_E$) is believed to result from the nuclear fusion reactions
of light elements.  The energy released in each of the chains producing
neutrinos is well known from laboratory experiments.  Consequently, if the
flux of the associated neutrinos can be determined then the photon
irradiance and luminosity can be inferred from those fluxes.

This is usually formulated \cite{bahcall2002} as:
\begin{equation} 
{L_\sun\over 4\pi\au^2} = I_E\quad {\rm and}\quad {L_\sun(\nu)\over
4\pi\au^2} 		= \sum_i(\alpha_i\Phi_i)
\end{equation} 

where $L_\sun$ is the total solar luminosity in photons, \au{} is the
average Earth-Sun distance, $I_E$ is the mean irradiance determined by
Earth-orbit satellites to be $1358.8\,{\rm W\,m^{-2}}
(8.482\times10^{11}\,{\rm MeV\,cm^{-2}\,s^{-1}})$ with a systematic
uncertainty of about $0.4\%$
\cite{bahcall2002,frohlich,bahcall-rmp}.  The $\alpha_i$’s are the 
coefficients giving the energy provided by and associated with the $i$-th
neutrino flux $\Phi_i$ and $L_\sun(\nu)$ is the total solar luminosity
inferred from the neutrino fluxes.

If the Sun operates as we presently think it does \cite{bahcall-rmp,
bahcall-astroj-supp, bahcall-astroj} then the ratio $L_\sun(\nu)/L_\sun$
should be unity.  Significant departure from that expectation would signal
the presence of different sources of energy within the Sun
\cite{bahcall1964,bahcall1969,bahcall1996}.  Another important point is 
that, from the reaction positions in the solar interior, the energy carried
by the photons and by the neutrinos reaches the Earth with a huge separation
in arrival times.  The neutrinos arrive directly in 8 minutes while the
thermal photon energy arrives from the solar plasma after approximately
$40\,000$ years \cite{fiorentini}.  Consequently, finding a disagreement
between $L_\sun$ and $L_\sun(\nu)$ would have significant implications for
environmental consequences in the long term.

Because the sum of \pp\ ($91.5\%$) and \be7 ($7.4\%$) neutrinos are expected
to be associated with $>98\%$ of the total flux, it follows that a precision
measurement of \phipp, either alone or together with \phibe, will provide
the major direct test of $L_\sun(\nu)/L_\sun$.  Currently this comparison is
only known to about $25\%$ \cite{bahcall2003,robertson}.

There are additional reasons to make a more precise determination of
$L_\sun(\nu)/L_\sun$: the fact that the average photon irradiance is very
well measured \cite{frohlich,waple,crom} has previously led to its use as a
constraint in global analyses of solar neutrino experimental data.  Used
first as a demonstration of possible flavor oscillations of neutrinos
\cite{spiro} and more recently, as additional and more precise solar and 
reactor neutrino experimental results have become available, as a powerful
constraint to aid in establishing best present knowledge of solar neutrino
mass-mixing parameters and individual fluxes \cite{bahcall2003, fogli2002,
fogli2006, bahcall-pena, bahcall-gonzalez, gonzalez}.

At the same time, the Standard Solar Models (SSM) \cite{bahcall-rmp,
bahcall-astroj-supp, bahcall-astroj} have continued to make significant
improvements so that quite precise predictions for the fluxes have been
made.  For example, \phipp{} is predicted to $\pm 1\%$ and \phibe{} to
$\pm9.5\%$ \cite{bahcall-astroj-supp}\footnote{Note 1: $8.5\%$ of the
$9.5\%$ is contributed from the experimental uncertainty in $S_{34}$ nuclear
cross-section factor; however, new data from the LUNA collaboration
\cite{confortola} suggests this contribution may be reduced to $2.5\%$.  
Private communication C.  Pe\~na-Garay.}.  When these predictions are
compared against the fluxes found from global fits to all of the existing
solar and reactor data \cite{homestake98} the levels of agreement differ
significantly depending on whether the photon luminosity is used as a
constraint or not.  For example, the ratios of global-fit fluxes to SSM
predictions are (at $1\sigma$): for \pp, $1.01 \pm0.02$ and for \be7
$1.03^{+0.24}_{-1.03}$ \emph{with the luminosity constraint} but are
$1.38^{+0.18}_{-0.25}$ and $0.13^{+0.41}_{-0.13}$, respectively,
\emph{without the constraint} and leads to the poor knowledge of 
$L_\sun(\nu)/L_\sun$ noted above
\cite{bahcall2003, bahcall-pena, bahcall-gonzalez, gonzalez, 
bahcall-2004-prl}.

The question of how precisely direct measurements, of either \phipp{} or
\phibe{} must be made in future experiments has been cogently addressed in 
an important paper by Bahcall and Pe\~na-Garay \cite{bahcall2003}.  Related
and more recent considerations also have been made by others
\cite{bahcall-astroj-supp,robertson}.  The level of precision required 
depends strongly upon the specific physics questions to be addressed.  In
all cases, the demands on experimental techniques are severe.  For example,
the authors of Ref.~\cite{bahcall2003} carried out simulations of global
analyses utilizing all present data plus inclusion of potential future \pp\
and \be7 experiments with assumed capability of precisions ranging from
$1\%$ to $30\%$ (at $1\sigma$).  They find that a \be7 result of $\pm 5\%$
could improve knowledge of $L_\sun(\nu)/L_\sun$ from $1.4^{+0.2}_{-0.3}$ to
$1.07\pm 0.13$.  Increased precision on \be7 alone would not yield further
improvement; while a $\pm 1\%$ on \pp\ could achieve a remarkable $0.99\pm
0.02$ on the luminosity comparison by neutrinos.  The authors note that a
result of this accuracy would be \emph{``a truly fundamental contribution to
our knowledge of stellar energy generation and place a $\pm 2\%$ bound on
all sources of energy other than low energy fusion of light elements (i.e.,
\pp\ and CNO chains)''.} The HERON detector, as shown in the present paper, 
is intended to be capable of reaching precisions on both fluxes commensurate
with these goals.

\subsection{The \pp-I vs \pp-II and LMA-MSW} 
The relative magnitude of \phipp{} versus \phibe{} is a particularly
relevant parameter, on the one hand, bearing on the accuracy of the SSM and,
on the other, as evidence for the MSW effect in the LMA.  It is valuable to
have an experiment which measures both \phipp{} and \phibe{} since several
systematic errors tend to cancel in the ratio $\phibe/\phipp$.  In the SSM
there is a very strong anti-correlation between the two fluxes with a
coefficient of $-0.79$ to $-0.81$ \cite{bahcall-astroj-supp}.  If the
$\he3(\he4,\gamma)\be7$ reaction of the \pp-II chain were the only
terminating branch, only one \pp\ and one \be7 neutrino would be produced in
each cycle.  Otherwise there would be two \pp\ and no \be7 neutrinos if
$\he3(\he3,2{\rm p})\he4$ of the \pp-I branch were the terminating reaction
of the full fusion cycle.  What the actual relative reaction rates are
depends on several not yet accurately known details within the Sun such as
elemental abundances, temperatures and density.  (Present versions of the
SSM predict a ratio of the two reaction rates as $0.174$ \cite{bahcall-pena}
which implies, prior to oscillation, a value of $0.080$ for
$\phibe/\phipp$.) An independent and precise measurement of these relative
rates would be an important contribution to the understanding of stellar
processes and would permit a refinement of the use of the SSM in global
analyses of neutrino data.  The physics of the MSW effect \cite{msw} is
embodied in the flavor-dependent interaction differences for neutrinos
propagating in matter as opposed to vacuum.  Due to the differences in
neutrino energies and solar density at their production points the \pp, \be7
and \B8 neutrinos are expected to have quite different survival
probabilities.  The LMA-MSW solution specifies what this energy dependence
must be.

The oscillations of the much higher energy \B8 neutrino should be strongly
suppressed by matter dominance and the \pp\ neutrinos much less since they
should be vacuum-dominated.  The \be7 and pep neutrinos, having energy
intermediate to \pp\ and \B8, are in the crucial energy region where the
transition between matter dominance and vacuum oscillations is to be
expected.  The \B8 flux is now very well measured by the Super Kamiokande
(SK) and Sudbury (SNO) experiments [in Ref.~\cite{homestake98}, see e.g.
Fukuda et~al.  and Ahmed et~al.]; however, direct experimental evidence for
this MSW transition is still lacking and could be provided by an experiment
such as HERON.

\subsection{Other areas of interest related to neutrino properties} 
The \phipp{} and \phibe{} measured in ES are, by necessity, the fluxes of
active neutrinos.  Consequently, in the measurement of $L_\sun(\nu)/L_\sun$
an alternative interpretation of a result consistent, within errors, to
unity can be taken as the establishment of a limit on the presence of
sterile neutrinos.

Due largely to the lack of precision experiments on these two major
low-energy fluxes, there has been some leeway in the recent analyses of the
solar and reactor data which allows for consideration of several
well-motivated proposals for ``new physics'' in the neutrino sector.  Among
these are possibilities for non-standard neutrino interactions with their
environments (NSI) \cite{bakref}.  Within this class of models the
additional effects to be expected are strongly constrained by existing data;
however, in some cases they should be most pronounced in the energy
dependence of the fluxes of solar neutrinos $<1\,{\rm MeV}$.  In these cases
the effect is qualitatively similar, but differs quantitatively, from that
to be expected from the LMA-MSW transition in the matter- to
vacuum-dominated neutrino energy regions.  Two examples of this class, are
models with flavor-non-conserving neutral currents \cite{friedland} or with
mass-varying-neutrinos (``MaVaNs'') \cite{fardon}; the latter inspired by
possible insights into understanding ``dark energy''.  Confirmed evidence
found for NSI would place our knowledge for the mass-mixing parameters in
doubt; alternatively, such precision measurements of the matter-vacuum
transition region would also serve to establish new limits on the existence
of NSI.

New measurements of low energy solar fluxes can play only a rather limited
role in improving on present knowledge of $\tan^2\theta_{12}$ and
$\sin^2\theta_{13}$ limits.  To be useful, the new data would need to be
folded into a comprehensive analysis with all the other data (solar,
atmospheric and reactor) which from present data have already established
impressive errors on the parameters
\cite{robertson,fogli2006,gonzalez,maltoni}.  The potential for improvement 
in these parameters by low energy ES or CC experiments has been subjected to
a detailed study by the authors of Ref.~\cite{bahcall2003} who conclude that
without new physics even $\pm1\%$ on \phibe{} would make a negligible
improvement on $\tan^2\theta_{12}$ and a \phipp{} result $<\pm3\%$ is
required to improve the error by more than $15\%$.  The authors' conclusions
on any improvements to be expected on $\sin^2\theta_{13}$ are similar.

\section{Detector} 
\subsection{Requirements} 
There are stringent requirements placed on any detector designed to achieve
these goals.  In order to be sensitive to all active neutrino flavors, the
detection reaction is that of the elastic scattering from atomic electrons
(ES) in the target: $\nu_{\electron,\mu,\tau} + \electron^- \rightarrow
\nu_{\electron,\mu,\tau} + \electron^-$.  The ES event signature is the 
occurrence of only a single, low energy recoiling electron in the detector
medium.  The recoil spectra are continuous from zero with the \pp\
decreasing monotonically to a maximum energy of $261\,{\rm keV}$ while the
\be7 spectrum is nearly flat up to $664\,{\rm keV}$ maximum (see points labeled input in Fig.~\ref{fig:spectra}).  These 
recoil spectra place a premium on achieving as low an energy threshold as
possible.  The most dangerous backgrounds are those which can be created by
the appearance of electrons from Compton recoils of gamma rays or
radioactive decays within the medium of the target.  These backgrounds need
to be mitigated by a combination of event signatures, use of low activity
materials, depth and shielding.  In order to achieve the desired few-percent
precision, very large statistics event samples are required and systematic
errors arising from effects due to analysis cuts (e.g., fiducial volume,
thresholds, \pp\ and \be7 event separation, calibrations) must be minimized.
Although these details emphasize the challenges to be faced in constructing
an operating detector, there are two mitigating factors: the ES
cross-section is very precisely known due to experiment and electro-weak
theory and the expected fluxes are high ($\phipp = 5.99\times 10^{10}$ and
$\phibe = 4.84\times 10^9\,{\rm cm}^{-2}{\rm s}^{-1}$
\cite{bahcall-astroj-supp}).  Typically, this will result in an ES event 
rate of $\sim 2\,{\rm events\cdot tonne}$-${\rm day}^{-1}$ which implies
that only a modest-size fiducial volume (e.g., $\sim 10\,{\rm tonne}$) is
needed for a high statistics experiment.

\subsection{HERON configuration and detection processes.} 
\begin{figure}\begin{center} 
\includegraphics[width=.8\columnwidth]{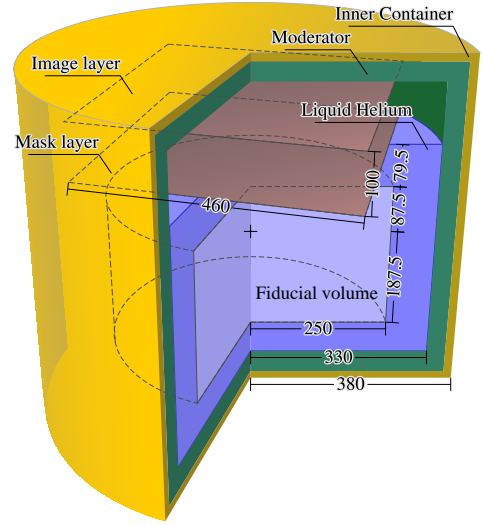} 
\caption{\label{fig:design}The geometry of the HERON detector design, all 
dimensions are in centimeters.}
\end{center} 
\end{figure} 

The detector design is shown in (Fig.~\ref{fig:design}) and its design is
discussed in more detail in Ref.~\cite{heron, adams, huang} .  The general
approach of HERON to these issues is as follows.  The target material chosen
is $\he4$ in the superfluid state (density 0.145 g/cc) which has several
beneficial properties.  Energy deposited in the helium by recoiling
particles can be detected by one or all of three processes: scintillation,
phonons/rotons or collecting the recoil electron trapped in a bubble.
Helium has no long-lived isotopes but more importantly it can be made
absolutely free of all other atomic species.  At superfluid temperatures it
is self-cleaning of impurities due to their high mobility and favorable
energy minimum at the container walls.  Even particulate matter quickly
attaches to the container walls at our operating temperature of $\sim
30$--$50\,$mK.  Since the bulk helium volume will be free of background
sources, the concern is to counter radiation entering from the cryostat
(27.3 tonne copper) and its environment.  In a separate study environmental
sources were modeled with the detector cryostat actively shielded externally
by $3.5\,{\rm m}$ of water at a rock overburden of $4500$
meter-water-equiv.~(m.w.e.).  Helium is virtually immune to creation of
long-lived cosmogenic muon activity, capture or decay in it; muons ($4\,{\rm
m}^{-2}{\rm day}^{-1}$ at 4500 m.w.e.) are vetoed externally and internally
in any case and greater depth is possible.  This was sufficient shielding
against environmental neutrons, cosmic muons and gammas that, for purposes
of the analysis simulation under consideration here, they would be
negligible relative to sources from the cryostat and other detector parts.
As a consequence the background issue reduces to controlling conversions of
gammas entering the helium volume of 21.6 and 8 tonnes total and fiducial,
respectively.  However, helium does not have good self-shielding properties
and this is partially compensated for by lining the cryostat with a
moderator of solid nitrogen enclosed in acrylic cells ($5.6\,{\rm T}$
acrylic and $104\,{\rm T}$ solid nitrogen, density $1.03\,{\rm g/cc}$).  The
function of the moderator is to absorb or degrade in energy by Compton
scattering the entering gamma rays which originate 97{\%} and 3{\%} from the
cryostat and moderator, respectively.  The flux of gammas entering the
helium is dominated by low energy (dominantly $<1\,{\rm MeV}$) cosmogenic
activity in the copper with the remainder from U, Th and other activity in
the copper, nitrogen, acrylic and other parts.  The background gamma
conversions in the helium cannot be fully eliminated but are amenable to the
development of a distinguishing signature which aids in their separation
from signal; the nature of this separation is detailed in Sec.~4 and 5.

We have chosen to use as the basic elements of detection for both signal and
background the collection of scintillation light and also the recoil
electron trapped in its bubble.  We have carried out studies of these
processes using prototype calorimeter sensors/detectors developed for this
project and suitable for both signal types; the processes and results are
discussed briefly below and in more detail in the references cited.

Excited or ionized He atoms along the electron path quickly form dimers in
the liquid.  The radiative decays to the ground states of these singlet and
triplet dimers emit photons in the ultraviolet.  The scintillation light is
in a narrow band centered at 16 eV and results from the decay of the singlet
dimer ${\rm He}_2^*(A^1\Sigma_{\rm u}^+)$ \cite{hillandketo}; since this
energy is lower than the first excited state of He at 20.6 eV, the liquid is
self-transparent.  $35\%$ of a recoil electron's energy is released ($\sim
{27000}$ photons/MeV) in this singlet dimer mechanism \cite{adams}; the
energy from the long-lived ($\sim 15\,{\rm sec}$) triplet dimer escapes or
is collision quenched.  (An additional $43\%$ of the recoil energy is
radiated in phonons and rotons which by quantum evaporation \cite{maris}
could in principle also be utilized with the calorimetric sensors for
discrimination \cite{bandler}; however, we find incorporating into event
signatures the detection of the recoil electron from the drifted bubble a
much stronger discriminant.)

When a recoil electron in He has lost most of its energy it forms an
electron bubble.  The electron experiences a strong, short-range repulsive
potential from the bound electrons on surrounding atoms due to Pauli
exclusion.  This repulsion forms a vacant volume, or bubble, of $19\,$\AA{}
radius in which the electron is confined.  The bubble forms in about 10
pico-sec with an effective displacement mass of $\sim 500$ He-masses and has
a hydrodynamic mass of half that \cite{fetter}; consequently due to this
difference in masses, under gravity and with no electric field, a bubble
would experience a buoyant acceleration of $\sim{2 g}$.  A uniform drift
velocity of the bubble can be provided and controlled by a combination of
applied electric field and a very low concentration of $\he3$; for example,
at $40\,$mK, $30\,$ppm \he3 and a field of $300\,$V/m provides a drift
velocity of $17.5\,$m/s.  In a worst case example of $5\,$m (maximum depth),
the $3\times10^8$ collisions induced due to the large cross-section of \he3
for scattering a bubble leads to an uncertainty in the transit time of
$16\,\mu$s and hence to a depth error of $< 1\,$mm.  In addition to
providing a ``drag'' force the \he3 also aids in extracting the electron
efficiently through the free surface of the liquid by vortex attachment
(Surko and Reif \cite{baskar}).  Two grids on either side of the liquid
surface (not shown in Fig.~\ref{fig:design}) provide the drift and
extraction fields
\cite{bhaskar}.  (At the normal operating temperature of $30$--$50\,$mK the 
vapor pressure is sufficiently low that the space above the liquid is
effectively a vacuum.) The final grid accelerates the electron to $\sim
1\,$keV thus providing a large and distinguishable pulse in the calorimeter.

For several reasons, photo-multiplier tubes (PMT) are not suitable event
detectors for HERON, among them: the high radioactivity of PMT's, poor He
self-shielding, lack of transparency of moderator and the desire to detect
the drifted electrons.  As mentioned, both scintillation and drifted
electrons are detected on the same calorimetric devices.  Each device
constitutes a pixel in a geometric array (a coded aperture) and consists of
a thin wafer of silicon or sapphire to which is attached a high sensitivity
metallic magnetic calorimeter (MMC) read out with a SQUID sensor
\cite{fleischmann}.  For astrophysical x-ray application, versions of MMC 
have been constructed with $\sim 3\,{\rm eV}$ resolution \cite{enss}.
Projecting from measurements on wafer prototypes of smaller heat capacity to
ones of the HERON size, $\sim 10\,$eV resolution is to be expected.  In a
full simulation of the response of this large wafer ($100\times 100 \times
0.4\,{\rm mm}$, in this example) with 16 eV photons it is found that single
photons should be detectable at wafer temperature of $40\,$mK producing a
pulse of $5\,$ms rise- and $100\,$ms fall-time \cite{kim}.  This performance
capability is assumed in the context of the analysis of Sec.~4 and 5.

For each neutrino event we must reconstruct its position within the He and
also its recoil electron energy.  In addition, we must develop event
signatures which aid in separating signal from background events.  The
maximum track length expected for a neutrino event is $\sim 2\,$cm so that
on the scale of the total helium volume ($149\,$m$^3$; $21.6\,$T) neutrino
signal events are effectively point sources of scintillation light.  At the
dominantly low energies of the gamma-ray background events the conversions
in He are overwhelmingly ($95\%$) Compton scatters.  $90\%$ of these
conversions are multiple depositions distributed over an average length of
more than $50\,$cm in the He.  Consequently, the scintillation from
background arises from a distributed, rather than a point, source;
additionally the event most often contains multiple, un-recombined electron
recoil bubbles.  These latter two features constitute the primary background
signature.  Subsequently, differences among the spectral and spatial
distributions of the signal and residual background events facilitate their
final separation.  In order to create these signatures and to enable the
necessary cuts on data samples the 2400 wafer calorimeters are arranged into
two planes in the vacuum space above the liquid; the resulting array
constitutes a coded aperture and provides the ability for both spatial and
energy reconstruction.

\section{Nature and use of HERON's coded aperture.} 
\subsection{General principles.} 
\begin{figure} 
\begin{center} 
\includegraphics[width=.8\columnwidth]{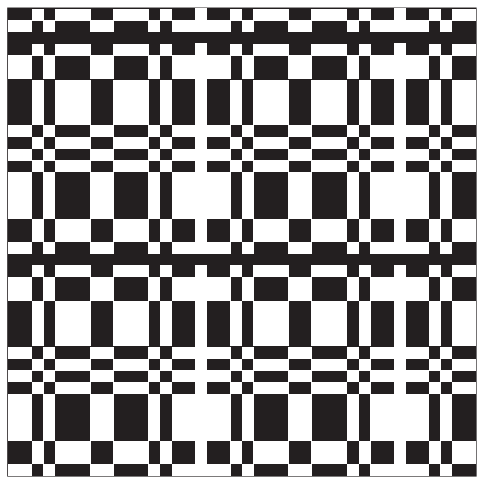} 
\caption{\label{fig:ura}URA pattern, $40\times 40$ in dimensions and 
$19\times 17$ in periodicity.  Dark and light areas represent opaque and
open regions, resp.  In the HERON mask, the smallest squares represents
single wafers.}
\end{center} 
\end{figure}

The concept of coded aperture arrays \cite {dicke,ura} has been a well
established one with arrays being widely used in x-ray astronomy .  The role
of the array is to accurately determine the \emph{direction} of incoming
photons from a remote x-ray point source.  A coded aperture array consists
of two parts, an imaging plane and, separated by a fixed distance, a mask
plane.  In the x-ray application the image plane consists of a set of active
sensors (or pixels) while the mask plane is opaque with a pattern of cut-out
apertures.  For far-distant sources, nearly parallel rays enter the array
and some of them are blocked by the opaque portions of the mask; thus the
image imposed on the sensor plane will resemble the pattern of the apertures
in the mask effectively leaving a ``shadow'' (See Fig.~\ref{fig:ura}).  In
principle, to determine the direction of the light source it is simply a
matter of comparing the shadow pattern to that of the mask itself.  In
practice, elegant techniques have been developed for design of mask patterns
and image deconvolution taking into account side-lobe effects as well as
intrinsic and statistical noise.  Various classes of mask patterns have been
employed in the x-ray field ranging from random apertures to strictly
repeating and regular patterns.  The choice of one over another depends upon
experimental considerations such as strength of photon flux, resolution
needed and sidelobe tolerance.

In the HERON application there are important differences: a) a non-distant
point source location is to be determined within a limited volume in
\emph{3-D}, b) the event energy must be measured, c) for reasons (a) and (b) 
both planes will consist of active wafer pixels and d) some discrimination
is needed between point and non-point sources.  The HERON coded aperture
array is arranged with the mask plane $1\,$cm above the liquid surface and
the image plane $1\,$m above the mask.  The pixels are arranged in
$40\times40$ square arrays ($460\times460\,$cm each) with $1600$ wafer
calorimeters in the image plane and $800$ in the mask.  The pattern of
apertures in the mask is that of a uniformly redundant array (URA)
\cite{ura}.  This mask pattern was chosen because of its low intrinsic noise 
and high transparency ($50\%$).  (In the notation of the URA it has a
(17,19) grid spacing \cite{ura} as shown in Fig.~\ref{fig:ura}.) The nature
of the URA and HERON physical properties and goals constrain the choice of
wafer pixel size.  The He fiducial volume can be chosen and varied during
physics analysis but the total mass of He is contained within a cylindrical
volume of $R=330\,$cm and $435\,$cm height.  In our application, the
transverse dimensions of the array should be at least commensurate with
those of the fiducial volume containing the sources and, although smaller
pixel sizes imply finer spatial resolution, ultimately photon statistics
dominate (typically a few hundred photons for many neutrino events, due to
energy and solid angle effects).  For the HERON geometry a choice of
$11.5\times11.5\,$cm pixels in a URA gives resolution adequate to the
physics goals without unnecessarily increasing the complexity or noise and
consistent with the desired single photon performance for a wafer of this
size.

\subsection{Loglikelihood method for event reconstruction.} 
The deconvolution techniques used for typical x-ray applications are not
applicable for our 3-D application; additionally, they are not easily
amenable to developing a background (distributed source) signature.
Instead we have adopted a likelihood approach along with a search algorithm
for the most probable position in 3-D space.

This approach treats each event in a sample containing both signal and
background as if it were a single point source.  With that assumption, it
finds from the observed photon hit pattern the most probable values of its
spatial location and total energy.  Although no attempt is made on an
event-by-event basis to distinguish signal from background, distributions of
the likelihood parameter's logarithm can be useful in separating signal and
background as we show in Sec.  5.  Similarly, the effective point-like
positions and effective energy distributions of the background events are
used.

Operationally, the algorithm initiates with a test-point location in the
volume and the probability of this test electron to produce the observed
photon hit pattern is calculated.  The test photon distribution is taken as
isotropic with straight-line propagation; the probability of hitting the
$i$-th wafer is then proportional to the solid angle ($\omega_i$) subtended
by the wafer from the photon current starting point.  After a systematic
search of points throughout the available volume, the location in space
found to have the highest probability is taken as the final position.

If $\Omega$ is the solid angle subtended by all $m$ wafers in both planes
and $N$ is the number of photon hits in the pattern then we can define a
quantity $N_{\rm tot} = 4\pi N/\Omega$ which is the total number inferred
for the test point.  Then the probability of an event located at $\mathbf x$
producing the recorded photon pattern is evaluated as:
\begin{eqnarray}\label{eq:likelihood} 
P(\mathbf{x}) = \left(1\over 4\pi\right)^{N_{\rm tot}}
\prod_{i=0}^m {\omega_i(\mathbf{x})^{n_i}\over n_i!} 
\end{eqnarray} 
where $n_i$ is the number hitting $i$-th wafer and $n_0\equiv N_{\rm tot}-N$
and for computational convenience we use the logarithm (loglikelihood):
\begin{eqnarray}\label{eq:loglikelihood} 
\mathcal{L}(\mathbf{x}) \equiv\ln P(\mathbf{x}) = -N_{\rm tot}\ln 4\pi 
	+ \sum^m_{i=0}\ln{\omega_i(\mathbf{x})^{n_i}\over n_i!}
\end{eqnarray} 
and select as the final position the one with the largest (least negative)
$\mathcal{L}$.  The final energy estimate is scaled from the solid angle
subtended from the test point.  The process converges rapidly guided by a
set of empirically established criteria for avoiding subsidiary maxima and
reaching a stable solution.\footnote{One might expect the determination of
the effective solid angle and efficiency to depend on the optical properties
of the UV radiation for He and the wafers.  Effects resulting from: the
width of the singlet dimer spectrum, internal reflection and refraction at
the He surface, Rayleigh scattering, multiple reflections between He and
wafers or among wafers themselves can affect the value of the effective
solid angle.  However, these effects are small due to the low refractive
index of He (1.045), the self-transparency and very long Rayleigh
mean-free-path (200 m
\cite {seidel}), the narrowness of the UV spectrum and the low reflectivity 
of the wafers at the incident angles involved \cite{reflect}.  These effects
can be taken into account by including ray-tracing techniques and we have
tested them in our analysis \cite{huang}; however, it is more
computationally demanding and results in no essential differences on flux as
we demonstrate in our final error table (Table~\ref{tab:fitting results}).
Consequently, for simplicity of analysis and discussion we ignore these
effects in this paper.}

\subsection{Event simulation and reconstruction} 
\begin{figure} 
\begin{center} 
\includegraphics{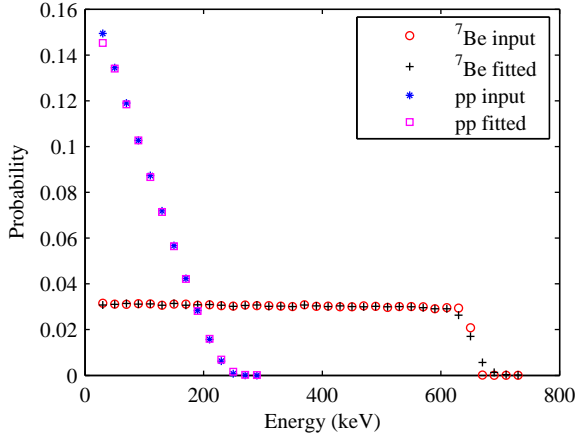} 
\caption{\label{fig:spectra}Comparison between input neutrino spectra and 
their reconstructed counterparts.  Each spectrum contains $\sim 5\times10^5$
simulated events and is normalized.}
\end{center} 
\end{figure}

A test of the reconstruction ability of the coded aperture approach has been
done in a way which examines a full range of variable correlations.  We have
generated samples of \pp, \Be7 (${5 \times10^5}$, each) and background
events (${2.5\times10^6}$) as they would appear in the configuration of
HERON described.  For the neutrinos, the input recoil electron energy
spectra are as shown in Fig.~\ref{fig:spectra} and the events are
distributed uniformly throughout the full He volume.  The input background
sample in the He is generated by propagating gamma rays initiating from
sources within the detector's principal components using GEANT3 \cite
{geant}.  The source activities and concentrations are listed in
Table~\ref{tab:bg sources}.  Within the He account was taken for
bremstrahlung, very low energy Compton recoils and delta-rays.  For all
input samples, the original position and deposited energy for every recoil
electron was retained for use in comparing to reconstructed values.  The
events were then reconstructed as described in Sec.~4.2.

\begin{table} 
\begin{center} 
\begin{tabular}{l | c | c | c} 
	&Cryostat	&Acrylic	&Solid N$_2$\\\hline ${}^{238}$U, ${}^{232}$Th
&$2\times10^{-13}\,$g/g	&$4\times10^{-13}\,$g/g	&$<10^{-16}\,$g/g\\
Cosmogenics	&$50\,\mu$Bq		&---			&---\\ ${}^{7}$Be, ${}^{39}$Ar, ${}^{85}$Kr
&---			&---			&$16\,$Bq/tonne\\ K$_{\rm nat}$	&$5\times10^{-10}\,$g/g
&$4\times10^{-9}\,$g/g	&---
\end{tabular} 
\caption{\label{tab:bg sources}Assumed levels of residual activity in 
major detector components.}
\end{center} 
\end{table} 

\emph {Signal reconstruction.} The principal interest in this section concerns the reconstruction results 
for the true point sources---the \pp\ and \Be7 events.

\begin{figure} 
\begin{center} 
\includegraphics{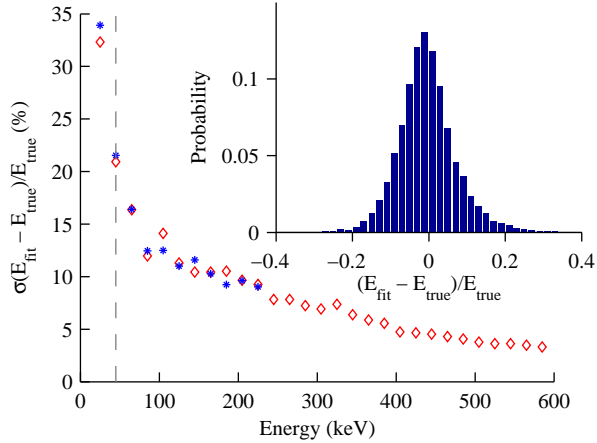} 
\caption{\label{fig:sigma_e}Resolution of reconstructed energy for neutrino 
events with respect to the energy of input events (asterisks: pp; diamonds:
Be).  The inset shows the distribution of reconstructed energy of events
between $100$ and $120\,$keV.}
\end{center} 
\end{figure}

\begin{figure} 
\begin{center} 
\includegraphics{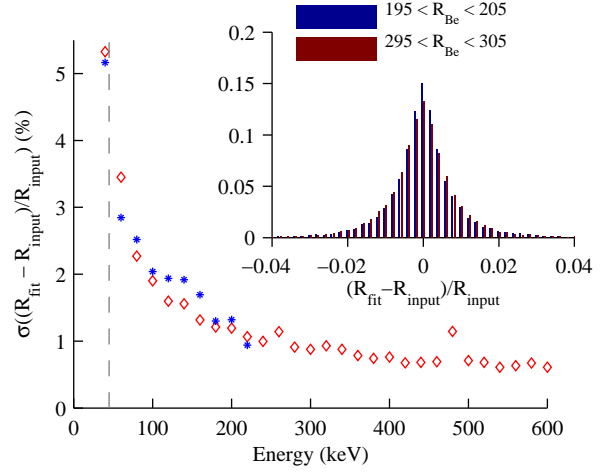} 
\caption{\label{fig:sigma_r}Resolution of reconstructed radial position 
for the neutrino events depending on their energies (asterisks: pp;
diamonds: Be) .  The inset figure shows the distribution of reconstructed
radii of events at different horizontal distances from the center of the
detector.}
\end{center} 
\end{figure}

Some examples of spatial and energy resolutions (1-$\sigma$) for the
reconstructed \pp\ and \Be7 samples are shown in Fig.~\ref{fig:spectra},
\ref{fig:sigma_e}, and \ref{fig:sigma_r}.  $R_{xy}$ (calculated from the 
fitted $x$ and $y$ coordinates) is the horizontal position in the He
cylindrical volume and $E$ is the total deposited energy.  The
$z$-coordinate (depth) is a factor $\sim 2\times$ poorer when reconstructed
by coded aperture; however, since the bubble drift time gives better than
$1\,$mm resolution the depth position by timing is used instead.  While the
results of the reconstruction are quite satisfactory, several points should
be noted about the figures and the nature of the results.  The resolution
dependence on photon statistics is common to both energy and position and is
dominant at the lowest energies, but it can also arise at higher energies as
a contribution due to diminished photon solid angle from the extremities of
the detector and mask geometry.  As the two inserts for
Fig.~\ref{fig:sigma_e} and \ref{fig:sigma_r} show, the maximum value of the
distributions indicate excellent agreement between input and output values,
the shapes near the maxima are symmetric and the FWHM look reasonable.
However, such figures are too limited to adequately illustrate the effects
of correlations nor of any asymmetries in the tails.  The correlations
between spatial and energy variables are of major concern as they enter into
flux accuracy when making the high level analysis cuts for fiducial volume
and threshold.  For our main purpose, the precise neutrino flux
determination, it is most direct and useful to examine comparisons of the
complete set of distributions to be used during the flux separation of the
three channels of events.  This is the case because if our model of the
detector is correct the full generation and reconstruction of these samples
naturally includes all of the effects of the correlations and tails and
permits estimation of their contributing errors.  Sec.~5 examines the
question of cuts to reduce background and to make flux separations.

\emph {Background reconstruction.} Any event trigger is accepted with a 
scintillation pulse above hardware threshold.  An important point in the
subsequent treatment of the events is dependent upon the detection of the
electron(s).  The difference in electron bubble topology (multiplicity)
between background and signal is an additional key factor in reducing the
raw background rate for analysis.

The nature and magnitude of the sources used in our simulation were listed
in Table~\ref{tab:bg sources}.  Any event giving one or more gamma
conversions anywhere in the entire helium volume is recorded and
reconstructed by the coded aperture method of Sec.  4.  The simulation
result implies a total daily rate in the full He volume of nearly $4 \times
10^3$ events; a factor $\sim 70$ greater than signal.  Several properties of
the background depositions are useful to establish prior to making high
level cuts on the samples for flux separation.  These properties
(multiplicity, event loglikelihood, energy and spatial distributions) were
determined from the sample of ${2.5\times10^6}$ simulated events.

\begin{figure} 
\begin{center} 
\includegraphics{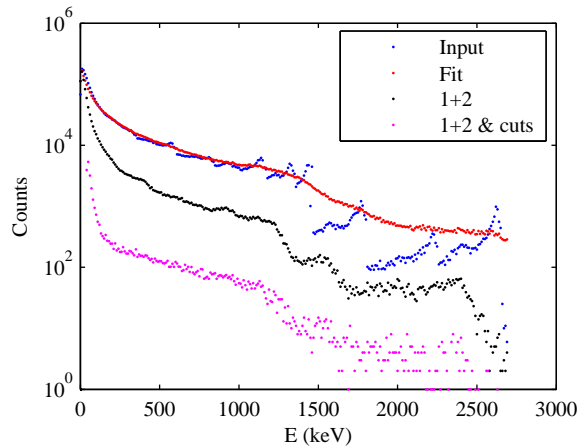} 
\caption{\label{fig:bg spectra}Energy spectra of simulated background events.  
``Input'' contains all events before reconstruction; ``Fit'' are the
reconstructed energy of these events; ``1+2'' are the events with no more
than 2 drifted electrons; ``1+2 \& cuts'' are the ``1+2'' with ``std''
fiducial and $45\,$keV energy threshold cuts.}
\end{center} 
\end{figure}

Due to the large size of the He volume and the low energies of the gammas
entering through the moderator, the detector is essentially hermetic and in
most cases fully contains the event.  Because of the He density and gamma
mean free path a typical event, prior to cuts, consists of widely
distributed Compton recoils with a mean multiplicity of 12.9 electrons and a
mean spatial separation of $57.6\,$cm.  The uncut energy deposition spectrum
both before and after reconstruction is shown in Fig.~\ref{fig:bg spectra}.
It should be emphasized that it is not important in our method to extract
the true energy of each background event, which in any case cannot be done
with a coded aperture for events of these topologies since there is no true
point-origin of scintillation.  In addition to providing a non-pointlike
$\mathcal{L}$-value, an effective energy is obtained and this accounts for
the smoothing seen in the reconstruction of the raw input spectrum for gamma
conversions in Fig.~\ref{fig:bg spectra}.

In order to use the number of detected drifted electrons effectively there
must be a high efficiency for collecting each one.  Full $100\%$ efficiency
is not possible due to ion re-combination.  As noted, for drift control
purposes we envision a uniform field of modest strength (300V/m).  With this
field and in our geometry very high efficiency ($\sim 85\%$) can be expected
for any electron with $> 5\,$keV \cite{bhaskar}.

As a first step in reducing the background sample for further analysis we
impose an event selection criteria of multiplicity $< 3$ detected electrons;
each with an energy $> 5\,$keV.  This reduces the background by a factor
$0.38$ without affecting the neutrino signal (possible bremstrahlungs from
neutrino events remain included by this cut).  See Fig.~\ref{fig:bg spectra}
for the resulting effect of this multiplicity cut on the spectrum.  Since
the goal is to extract pp and Be fluxes separately, we choose not to make an
event-by-event division of neutrino events from background with the
likelihood ratio.  Instead we defer to making the simultaneous separation of
all three channels by the method described in Sec.  5.  Additional samples
of ${8\times10^5}$ background events with detected multiplicity $< 3$ are
also generated. These, together with the neutrino samples, are then binned
to form PDF's and used for further study of high level cuts.

\section{Flux separation\label{sec:flux sep method}} 

Fundamental to the method of flux separation used here is the construction
of accurate probability distribution functions (PDF).  They present the
expected appearance for four major reconstructed variables  ($R^2$, $z$, 
energy and ${\cal L}(\mathbf{x})$) for events from
the two neutrino signal channels and the backgrounds.  The PDF's are used in
an extended-maximum-likelihood method \cite{lyons} applied to a simulated
experimental random sample including events from all three channels.  The
PDF variables used are the radial and vertical positions of the events,
their energies and the $\mathcal{L}$ of Eqn.~\ref{eq:loglikelihood}.  Our
principal goal here is to examine the type of systematic errors introduced
by this approach.

\subsection{Fitting fluxes using likelihood method\label{sec:fitting method}} 
After we have observed and reconstructed neutrino and background events with
sufficient statistics, we may separate the three fluxes (pp, \be7 and
background) using the extended likelihood method.  Given certain binning,
the extended (log)likelihood function is defined for each individual variable
as
\begin{equation} 
{\mathcal{L}_\Phi} = - \sum_{\alpha=1}^3 N_\alpha + \sum_{{\rm bin}\ i} n_i
	\ln \sum_{\alpha=1}^3 N_\alpha p_{\alpha i} 
\end{equation} 
where $\alpha$ denotes the flux type, $p_{\alpha i}$ the probability of flux
$\alpha$ populating the bin $i$, $n_i$ the total number of events in bin
$i$, and $N_\alpha$ the number of events being of type $\alpha$.
$N_\alpha$ are the parameters that are to be adjusted to maximize
$\mathcal{L}_\Phi$, with the constraint that their sum equals to the
observed total number of events.

The derivation and structure of error $\epsilon_\alpha$ for each fitted flux
$N_\alpha$ by this method is discussed in (\cite{huang} Appendix A).  The
error matrix $V$, defined as
$V_{\alpha\beta}\equiv\left\langle\epsilon_\alpha
\epsilon_\beta\right\rangle$, in this context evaluates to 
\begin{equation}\label{eq:error matrix} 
V = -{\mathbf N} + {\mathbf P}^{-1},
\end{equation} 
where elements of matrices $\mathbf N$ and $\mathbf P$ are
\begin{equation}\label{eq:N and P} 
{\mathbf N}_{\alpha\beta} \equiv \delta_{\alpha\beta}N_\alpha
\quad{\rm and}\quad 
{\mathbf P}_{\alpha\beta} \equiv \sum_{{\rm bin}\ i} {p_{\alpha i} p_{\beta
i}
\over n_i},
\end{equation} 
where $\delta_{\alpha\beta}$ is the Kronecker delta, with indices $\alpha$ and 
$\beta$ referring to flux types.

Note that the elements in matrix $\mathbf P$ may have very large values if
some bins have very low counts (i.e.  $n_i$ being small), which will lead to
reduced numerical stability of ${\mathbf P}^{-1}$ and consequently of $V$.
To avoid this, we chose to use a variable-width binning where bin edges are
chosen so that each bin contains roughly the same number of events.

Matrix $\mathbf P$ describes how much one PDF resembles another.  If for two
fluxes $\alpha$ and $\beta$, $p_{\alpha i}$ and $p_{\beta i}$ are strongly
correlated over index $i$, $\mathbf P$ will have large non-diagonal
elements, resulting in large valued elements in ${\mathbf P}^{-1}$ and $V$.
In other words, if two PDFs are similar in shape, the expected errors of
their fitted fluxes will be high.  This property dictates the variables we
can use for flux separation: reconstructed $R$ and $z$ distributions are
basically identical for both types of neutrinos, thus they are not suitable
for separating \be7 from \pp; while energy and loglikelihood of
reconstructed events provide satisfactory PDFs for this purpose.

\subsection{The PDF distributions.} 
\begin{figure} 
\begin{center} 
\includegraphics{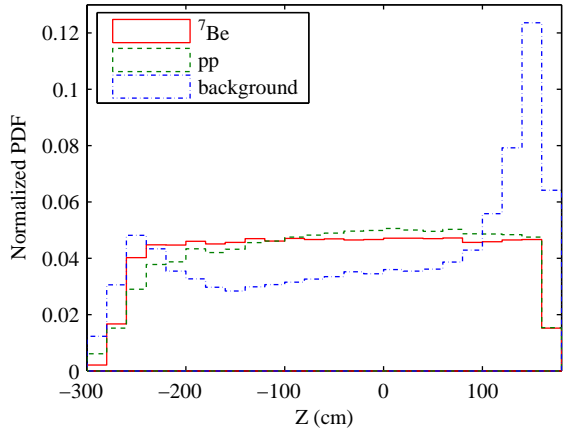} 
\caption{\label{fig:Z hist}Normalized PDF distribution of depth of simulated events \emph {before high level cuts}.  See 
Fig.~\ref{fig:design} for the coordinate system.}
\end{center} 
\end{figure}

 In generating the PDF's, the input \pp\ and \Be7 neutrino oscillated fluxes
to be expected at Earth are based on the Bahcall-Pinsonneault solar fluxes
\cite{bahcall-astroj-supp, bahcall-astroj}, the mass and mixing parameters 
from the global fit of Fogli et al \cite{fogli2006} and assumed two-flavor
mixing.  ${5 \times10^5}$ recoil electron events were generated for each of
\pp\ and \Be7 and each contained the appropriate mix of active neutrino 
flavors ($\nu_{\rm e}$, $\nu_{\mu\tau}$).  (The resulting daily event rate
above 45 keV is 1.7 \pp\ and 0.6 \Be7 per tonne.) The points labeled
``input'' in Fig.~\ref{fig:spectra} show the spectra in HERON to be
reconstructed.  For background $2\times10^8$ gamma rays due to the sources
listed in Table~\ref{tab:bg sources} were generated and propagated via GEANT
through the detector.  Of these, a sub-sample of $8\times 10^5$ with less
than three conversion electrons in the He each with energies greater than
$5\,$keV were used in creating the background PDFs.  This sub-sample choice
resulted from the detailed study of the properties of a full sample of
conversions as outlined in Section 4.3.  Shown in Figures \ref{fig:E
hist}--\ref{fig:Z hist} are samples of the four PDF's prior to any fiducial
or threshold cuts.

\begin{figure} 
\begin{center} 
\includegraphics{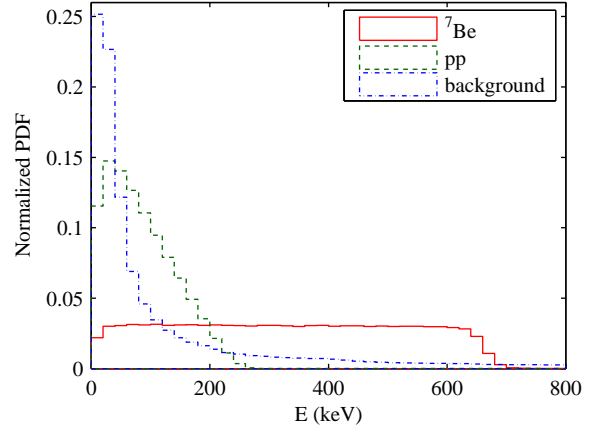} 
\caption{\label{fig:E hist}Normalized PDF for reconstructed energy spectra of simulated events \emph {before high level cuts}.  
These include ${8\times10^5}$ background events, and ${> 5\times10^5}$
events in each neutrino flux.}
\end{center} 
\end{figure} 

\begin{figure} 
\begin{center} 
\includegraphics{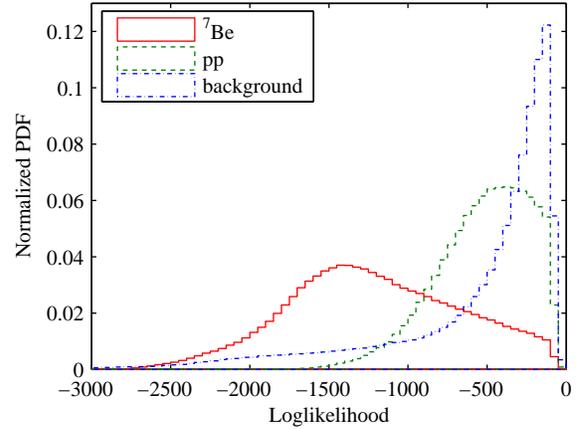} 
\caption{\label{fig:LL hist}Normalized PDF distribution of fitted loglikelihood of simulated 
events \emph {before high level cuts}.}
\end{center} 
\end{figure} 

\begin{figure} 
\begin{center} 
\includegraphics{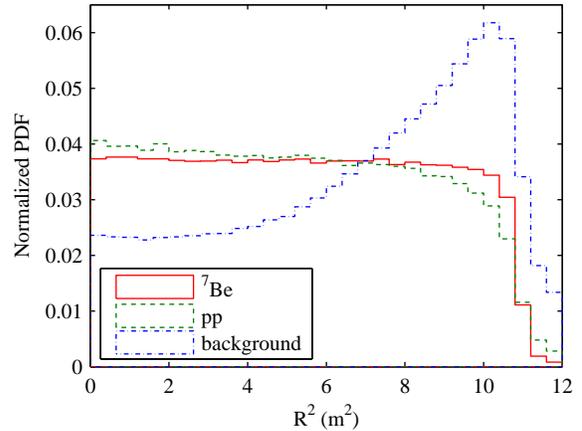} 
\caption{\label{fig:R hist}Normalized PDF distribution of horizontal radii $R=\sqrt{x^2+y^2}$ 
for the simulated events \emph {before high level cuts}.}
\end{center} 
\end{figure} 

\subsection{High-level cuts.} 
Even after reducing background rate by rejecting events with three or more
recoil electrons, the background conversion rate in the full helium volume
would still be a factor $\sim 25$ higher than that of the neutrino recoils
in the full helium volume, this factor can be significantly improved by
instituting fiducial volume and energy threshold cuts.

 The following considerations were made for choosing to impose these high
level cuts on fiducial volume and threshold with a minimum of bias.
Depending on the fraction of detector surface area covered by the coded
aperture, the resolution worsens somewhat at the extreme edges.  In practice
the position and energy resolution will have to be carefully calibrated by
temporary insertion of sources at measured locations; however, cutting in
the simulation avoids possible artificially introduced features in the PDF's
due to inaccuracies in the reconstruction model.  Additionally, it is convenient
to work with a better signal to noise ratio provided that the distinguishing
characteristics among the PDF's are not compromising the sensitivity.  A
check on this can be seen in Table 2 and 3 where the results are presented
for analyses with three quite different fiducial volumes and two differing
thresholds.  The contribution to flux error is not significantly changed
amongst them.  Our goal has been to seek an optimal set of cutting
parameters that yield the lowest relative systematic uncertainty in the
fitted neutrino fluxes.

The various components of errors due to the flux separation process and the
high-level cuts are discussed in the next section.  For this purpose we have
considered a sample size equivalent to 5 years of running in the HERON
design.  This would result, using the ``standard'' fiducial volume (see
below), in ${2.7\times10^4}$ \pp, ${9.3\times10^3}$ Be and $7.5\times10^4$
background events.  However, in examining cut effects we studied six
combinations of energy threshold cuts and fiducial cuts using the new PDF's
appropriate to the cut combinations involved.  The energy thresholds tested
are at $30$ and $45\,$keV, and the fiducial cuts paired with them are
referred to as ``small'' ($-177.5\,{\rm cm} < z < 77.5\,{\rm cm}$,
$R_{xy}<230\,$cm), ``std'' ($-187.5\,{\rm cm} < z < 87.5\,{\rm cm}$,
$R_{xy}<250\,$cm), and ``big'' ($-208.5\,{\rm cm} < z < 108.5\,{\rm cm}$,
$R_{xy}<270\,$cm), whose coordinate system is shown in
Fig.~\ref{fig:design}.  Interestingly, even though the S/N ratios can be
very different depending on the cuts, the overall errors of fitted neutrino
fluxes are much less sensitive.

\subsection{Errors in flux separation.\label{sec:error in fluxes}} 
Besides the statistical error on the total sample there are systematic
errors in the flux separation process: error due to volume and threshold
cuts as a result of spatial and energy resolution of the reconstruction,
error from intrinsic uncertainty of the likelihood fitting method (arising
as discussed in Sec.~\ref{sec:fitting method}) and error due to uncertainty
of absolute energy scale.  We treat the latter error separately from the
others.

Error due to volume and threshold cuts stems from the inaccuracy of spatial
reconstruction of events.  Given a set of $N$ events, one can expect some
$N_a$ of them to be mistakenly accepted, while some other $N_r$ of them to
be mistakenly rejected.  $N_a$ and $N_r$ can be modeled by comparing
reconstructed events with their corresponding input values from our
simulated event data, and the difference $N_a - N_r$ can be considered as a
form of systematic bias of fitted flux, which could in principle be
corrected for but the correction would be very small and of the same order
as its error.  Both $N_a$ and $N_r$ are subject to statistical fluctuations
and such a correction will have an uncertainty $\sigma=\sqrt{N_a + N_r}$.
This error depends on both spatial and energy cuts, and its values for some
of their combinations are listed in Table.~\ref{tab:volume cut error}.
\begin{table} 
\begin{center} 
\begin{tabular}{l | c r | r r r | r} 
	&Volume	&$E_{\rm thresh}$ 	&Flux 	&$N_a$ 	&$N_r$ 	&$\sigma/N$\\\hline 	&big
&$30\,$keV 	&$12899$ 	&$345$ 	&$256$ 	&$0.19\%$\\
\be7	&std	&$45\,$keV 	&$9345$ 	&$236$ 	&$169$ 	&$0.22\%$\\ 
	&small	&$45\,$keV	&$7339$		&$184$	&$127$	&$0.24\%$\\\hline 	&big &$30\,$keV
&$42123$	&$2256$	&$2037$	&$0.16\%$\\ pp	&std 	&$45\,$keV &$27000$ 	&$1335$
&$1233$	&$0.19\%$\\ 	&small	&$45\,$keV	&$21163$	&$1025$ &$963$	&$0.21\%$\\
\end{tabular} 
\caption{\label{tab:volume cut error}Flux error due to volume and threshold cuts, assuming 
event observed over a period 5 years. The definitions of $N_a$, $N_r$ and volume sizes are 
given in text.}
\end{center} 
\end{table} 

Errors of separated fluxes inherent to the likelihood method, as discussed
in Sect.  5.1, are also calculated based on simulated neutrino and gamma
event data.  Since these errors depend on the shapes of the PDFs, which in
turn depend on fiducial and energy cuts, they are listed in
Table~\ref{tab:fitting error} in relation to the cuts also.  The
distributions of fitted energy are used to provide PDFs in the table.
\begin{table} 
\begin{center} 
\begin{tabular}{l r | r r r} 
Volume	&$E_{\rm thresh}$ 	&$\varepsilon_{\rm Be}/N_{\rm Be}$
&$\varepsilon_{\rm pp}/N_{\rm pp}$ 	&$\varepsilon_\nu/N_\nu$\\\hline big
&$30\,$keV 	&$2.48\%$ 	&$1.01\%$ 	&$1.00\%$\\ std	&$45\,$keV 	&$2.48\%$
&$1.06\%$ 	&$1.08\%$\\ small	&$45\,$keV 	&$2.56\%$ 	&$1.11\%$
&$1.10\%$\\\hline
\end{tabular} 
\caption{Error inherent to the likelihood method using PDFs.  
$\varepsilon$s are obtained from the error matrix $V$.  $\varepsilon_\nu$
and $N_\nu$ are the error and total neutrino flux if the \pp\ and \be7 are
not fitted separately.  5 years of observation is assumed.\label{tab:fitting
error}}
\end{center} 
\end{table} 

It is worth noting that we are using only the PDF of one of the several
fitted variables available.  In principle, for example, we could fit the
fluxes from both energy and ${\mathcal{L}(\mathbf{x})}$ value PDFs, which, when
combined, could lead to smaller fitting errors.  However, there are
significant correlations in fitted fluxes from energy and
${\mathcal{L}(\mathbf{x})}$, thus the combined fitting does not yield significant
improvement over fitting using either variable separately.  For simplicity, we currently choose to fit using only the
energy PDFs, which have slightly smaller errors than those from the
${\mathcal{L}(\mathbf{x})}$.

Error induced by absolute energy scale uncertainty is a consequence of any
energy threshold cut, because for events near the energy threshold, an error
in energy scale may change its status of being accepted or rejected.  This
type of error is commonly described by the quantity ${\rm d}\Phi/\Phi\over
{\rm d}E/E$, which, when multiplied by the energy scale uncertainty $\delta
E/E$, gives the relative error in flux $\Phi$.  Since the absolute energy
scale uncertainty is a property of the physical detector system, it will be
obtained by insertion of a calibration source.  For purposes of this
analysis it is assumed that a $2\%$ uncertainty on the scale is achievable
leading to a contribution to the pp flux error of $0.94\%$ at a 45keV
threshold.  Based on our simulated fluxes, values for this quantity are
listed in Table.~\ref{tab:d_phi_d_E}.  Note that this error depends on the
energy threshold value but not the fiducial cut parameters, because the
neutrino events are uniformly distributed throughout the detector volume.
The calibration source would also be needed to establish the accuracy of the
position reconstruction.
\begin{table} 
\begin{center} 
\begin{tabular}{l | c c | c c } 
&\multicolumn{2}{c|}{\be7}&\multicolumn{2}{c}{pp}\\\hline $E_{\rm thresh}$
&$30$ 	&$45$	&$30$	&$45$\\\hline ${\rm d}\Phi/\Phi\over {\rm d}E/E$	&$0.056$
&$0.073$	&$0.29$ &$0.47$\\
\end{tabular} 
\caption{\label{tab:d_phi_d_E}Coefficient of error from absolute energy scale uncertainty.} 
\end{center} 
\end{table} 

These three types of errors are all considered as systematic errors, and are
independent of each other, thus will be summed in quadrature.
Table~\ref{tab:fitting results} lists the combined error and the statistical
fluctuations of the fitted fluxes (without the absolute energy scale
uncertainty part, which should add another $\sim 1\%$ to pp and combined
neutrino fluxes if we assume $2\%$ absolute energy scale uncertainty).  5
years worth of observation at ``std'' fiducial cut and a $45\,$keV energy
threshold are used in that table.
\begin{table} 
\begin{center} 
\begin{tabular}{l | c c c | c c c | c c} 
&	\multicolumn{3}{c|}{Statistical} 	&\multicolumn{3}{c|}{Systematic}
&\multicolumn{2}{c}{$\sigma_{\rm scale}$}\\ &\be7	&pp 	&\be7+pp &\be7	&pp
&\be7+pp &\be7	&pp\\\hline
\mbox{non-opti}& $1.03\%$	&$0.61\%$ 	&$0.52\%$ 
		&$2.49\%$ 	&$1.08\%$ 	&$1.10\%$ 		&$0.15\%$ 	&$0.94\%$\\ optical &$1.03\%$
&$0.60\%$ 	&$0.52\%$ 		&$2.78\%$ 	&$1.21\%$ 	&$1.19\%$ &$\sim0.2\%$ 	&$\sim
1\%$\\
\end{tabular} 
\caption{\label{tab:fitting results}Relative statistical and systematic 
error of fitted neutrino fluxes.  $\sigma_{\rm scale}$ column lists the
error from absolute energy scale uncertainty of $2\%$.  The terms
``optical'' and ``non-optical'' are defined in footnote 2.}
\end{center} 
\end{table} 
For comparison, we also show in Table~\ref{tab:fitting results} the
resulting errors to be expected when all optical effects are included.  The
differences are not significant as we have discussed in footnote 2.

\section{Summary and Conclusions} 

We have described that, on the basis of the new knowledge gained in recent
years of neutrino properties and of higher energy solar neutrino fluxes,
there are excellent reasons to perform precision real-time measurements of
the very low-energy neutrino fluxes from the Sun.  The physics goals
outlined in Sec.  2 include determining the luminosity of the Sun in
neutrinos, providing checks on some details of the SSM, testing the MSW
effect in the LMA solution and improving constraints on the neutrinos
mass-mixing parameters as well as providing discovery opportunities in the
new low energy regime.

To achieve these goals detectors are required which can measure the \phipp\
flux with a precision better than $3\%$ and the \phibe\ flux to better than
$5\%$.  Such detectors must be capable of collecting very large event
samples and maintain good control of systematic errors.  We have described
the design of such a possible detector, HERON, and have simulated its
performance.  Although the HERON detector is not presently scheduled or
funded for construction, by experimentation in prototypes of several liters
we have measured the details of energy loss processes (scintillation,
phonons/rotons, electron bubbles) for low energy electrons in the superfluid
\cite{adams,huang,bandler,baskar}.  The development
of wafer calorimeters capable of detecting all three channels has been
carried out \cite{fleischmann,enss,bhaskar,kim}.  We have not tested an array
of wafers as a coded aperture; however, given the well tested use of the
method in other fields, performance as simulated can be reasonably expected.
The superfluid helium target material is itself free of intrinsic internal
background and provides two channels (scintillation and drifted electrons)
to distinguish and separate externally entering background from point-like
neutrino ES signals via an active coded-aperture array.

The simulation has been directed towards establishing the systematic and
other errors to be expected for the HERON detector in an
exposure of 5 years.  For that purpose large samples of both signal and
background events were generated and then fully reconstructed according to
the physical processes in helium, the detector geometry and the properties
of the coded aperture design.  The expected signals were based on current
best understanding of the neutrino mass-mixing parameters and the well-known
electroweak scattering cross-sections.  The backgrounds (gamma-ray
conversions in the helium) were simulated from radioactive sources
distributed throughout the major materials surrounding the helium.  The
level and nature of the activities assumed was in line with current best
practice in solar and double beta decay neutrino experimentation.  To
separate a combined sample of pp, Be7 and backgrounds into their respective
flux components, an extended loglikelihood method was used employing
probability distribution functions constructed from various samples of the
above simulation prescription.  By design, the method included all
correlations among variables imposed by event reconstruction, various cuts
on the data as well as those arising from the properties of the PDF's
themselves.

The results are quite promising as can be seen in Tables 2-5.  It appears
that should the detector be built and perform as modeled it would be capable
of satisfying the criteria necessary for the precision pp and Be7 flux
measurements.  For example, to take the particular choice of the so-called
``standard threshold and fiducial cut'' on a 5-year exposure and combining
all errors except energy scale (statistical, high-level cuts, likelihood
method) a precision on pp flux of ${\pm 1.35\%}$ (or ${\pm 1.68 \%}$
including energy scale uncertainty) results.  Similarly for \phibe, we find
errors of ${\pm 2.96\%}$ and ${\pm 2.97\%}$ without and with, respectively,
the energy scale uncertainty included.  The full neutrino flux obtained
without attempting to employ the separation of \phipp{} and \phibe\
individually would present a combined error of ${\pm 1.31\%}$.

The validity of the background model used in these simulations is a key
issue.  The composition and relative magnitude of the background sources
assumed were based on current experience in the field and should therefore
be realizable in practice; nonetheless, it is important to ask to what extent are the
simulation results dependent upon the assumed model and to what extent can
the model be checked in practice.  The decay modes of the sources, branching
ratios and energies of the decay products are well known and the method of
their propagation through the simulation programs are well established.
Perhaps most important is the question of whether we may have mis-estimated
the total background level and if so how strongly the result would be
affected.  A mistake in the magnitude would enter principally through the
separation procedure using the PDF's; consequently we have tested this
effect by varying the assumed size of the background over a wide range.  The
effect is not drastic; for example, should the background be 5 times larger,
the
\phipp\ error would double while a factor 50 larger background would raise 
the \phipp\ error by six times (however, this latter rate would introduce
prohibitive wafer deadtime).  In contrast, reducing the background by a
factor of $0.5$ only improves the \phipp\ error by $15\%$.

In practice, there are some checks available on the model.  Due to the good
position and energy resolution for the point-like signal events, fiducial
volumes of various sizes can be made and the stability of the flux results
checked.  As we have seen in Tables 2, 3 and 5, varying the high-level cuts
by setting different fiducial volumes and thresholds does not have a strong
effect upon the expected errors.  Similarly, the nature and dependence of the
observed spatial and energy distributions as a function of these cuts can
also be compared directly to the model.

In conclusion we believe that a detector of the HERON design utilizing
superfluid helium and a coded aperture array could provide the capability to
carry out the multiple physics goals achievable through precision,
real-time, simultaneous measurements of \phipp{} and \phibe.

\section{Acknowledgements} 
We are grateful to the U.S.  Department of Energy for support of R\&D on
this project through grant DE-FG02-8840452.  We are indebted to J.R.  Klein
for his close reading and valuable suggestions, R.B.  Vogelaar for comments
and A.W.  Poon for assistance in providing additional computing resources at
Lawrence Berkeley National Laboratory.

\appendix 
\section{Solar orbit eccentricity.} 
In order to examine the effectiveness of our flux fitting method under a
more realistic context, we chose to test for the annual solar neutrino flux
oscillation due to the Earth orbit eccentricity.  The orbit of the Earth
around the Sun has an eccentricity of $\sim 1.67\%$.  Since the diameter of
the Sun ($\sim 1.4\times 10^6\,$km) is much smaller than the radius of the
orbit ($\sim 1.5\times 10^8\,$km), the Sun can be treated as a point source,
thus the neutrino flux observed on Earth will oscillate according to
$1/r^2$.

We simulated the number of events observed daily over a span of five years.
These event numbers consist of both \pp\ and \be7 events including their
statistical fluctuations.  A random error according to the systematic
uncertainty of the flux fitting method is then added to each day's flux, to
simulate the errors introduced during the reconstruction and flux separation
process.  Then the daily fitted event counts are grouped into consecutive
$60$ day periods, with $5$ days' worth of data each year discarded for
simplicity.  Thus over 5 years, each of these periods contain $300$ days'
worth of events.  A $\chi^2$ fit of these event numbers against the model of
an elliptic Earth orbit can then be performed, using the eccentricity as the
fitting parameter.

\begin{figure} 
\begin{center} 
\includegraphics{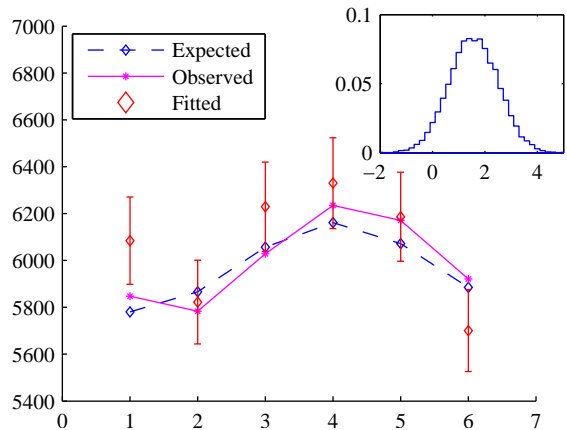} 
\caption{\label{fig:ecc}A typical fitting for Earth orbit eccentricity.  (Inset: Normalized eccentricity results from $2\times 10^4$ trial simulations.) } 
\end{center} 
\end{figure} 

Figure \ref{fig:ecc} shows a typical set of data for such fitting.  The
error bars on the ``fitted'' data include both statistical and systematic
errors, as discussed in Sec.~\ref{sec:error in fluxes}, where the absolute
energy scale uncertainty is taken as $2\%$.  Best fit of the particular data
set in that figure gives an eccentricity of $2.1\%$; repeating this
simulation for $2\times 10^4$ times reveals the distribution of the best
fitted eccentricity values to be $1.68\%\pm 0.95\% (1\sigma)$.  This
exercise demonstrates that our flux fitting method is capable of resolving
the solar neutrino flux to the precision of a few percent given 5 years of
detector running time.


\begin{thebibliography}{99} 
\bibitem{heron} 
R.E.~Lanou, H.J.~Maris, G.M.~Seidel, \emph{Phys.  Rev.  Lett.}
\textbf{58} (1987) 2498; S.R.  Bandler, et~al., \emph{J.  Low Temp Phys.} 
\textbf{93}~(3/4) (1993) 715; R.E.  Lanou, \emph{Nucl.  Phys.  B.  (Proc.  
Suppl.)} \textbf{138} (2005) 98.

\bibitem{dicke}R.H.~Dicke, \emph{Astrophys.  J.} \textbf{153} (1968) L101; 
J.G.~Ables, \emph{Proc.  Astro.  Soc.  Aust.} \textbf{4} (1968) 172.

\bibitem{ura} 
K.E.~Fenimore, T.M.~Cannon, \emph{Applied Optics} \textbf{17} (1978) 337.

\bibitem{bethe} 
H.A.~Bethe, \emph{Phys.  Rev.  C} \textbf{55} (1939) 248; J.N.~Bahcall,
\emph{``Neutrino Astrophysics''}, Cambridge University Press (1989) 

\bibitem{borexino} 
C.~Arpesella et~al, \emph{arXiv:0708.2251v1 [astro-ph]}; G.~Alimonti,
et~al.,
\emph{Astroparticle Physics} \textbf{16} (2002) 205--234.  


\bibitem{exps} 
D.N.~McKinsey, J.M.~Doyle (CLEAN; neon using ES), \emph{J.  Low Temp Phys.}
\textbf{118} (2000) 153--165; Y.~Suzuki (XMASS; xenon using ES), in: \emph{Proc.  
LowNu2}, World Scientific Publ., 2001, p.~81; R.~Raghavan (LENS; indium for
$\nu_{\rm e}$ flux), \emph{Phys.  Rev.  Lett.} \textbf{78} (1997) 3618;
H.~Ejiri et~al.  (MOON; Mo for $\nu_{\rm e}$ flux), \emph{Phys.  Rev.
Lett.}
\textbf{85}~(14) (2000) 2917; C.  Amsler et~al.  (CF4 TPC for ES) \emph{arXiv:0710.1049v1 [hep-ex]}; M.~Chen (SNO+; liquid scintilator for pep \& CNO 
fluxes), \emph{Earth, Moon and Planets} \textbf{99} (2006) 221.

\bibitem{msw} 
L.~Wolfenstein, \emph{Phys.  Rev.  D} \textbf{17} (1978) 2369;
S.P.~Mikheyev, A.Y.~Smirnov, \emph{Nuovo Cimento} \textbf{C9} (1986) 17.

\bibitem{bahcall2002} 
J.N.~Bahcall, \emph{Phys.  Rev.  C} \textbf{65}~(2) (2002) 025801.

\bibitem{frohlich} 
C.~Frohlich and J.~Lean, \emph{Geophys.  Res.  Lett.} \textbf{25} (1998)
4377.

\bibitem{bahcall-rmp} 
J.N.~Bahcall, M.H.~Pinsonneault and G.J.~Wasserburg, \emph{Rev.  Mod.
Phys.}
\textbf{67}~(4) (1995) 781--808.  

\bibitem{bahcall-astroj-supp} 
J.N.~Bahcall, A.M.~Serenelli and S.~Basu, \emph{Astrophys.  J.  Supplement
Series} \textbf{165} (2006) 400.

\bibitem{bahcall-astroj} 
J.N.~Bahcall, M.~H.  Pinsonneault and S.~Basu, \emph{Astrophys.  J.}
\textbf{555} (2001) 990.  

\bibitem{bahcall1964} 
J.N.~Bahcall, \emph{Phys.  Rev.  Lett.} \textbf{12}~(11) (1964) 300--302.

\bibitem{bahcall1969} 
J.N.~Bahcall, \emph{Phys.  Rev.  Lett.} \textbf{23}~(5) (1969) 251--254.

\bibitem{bahcall1996} 
J.N.~Bahcall, M.~Fukugita and P.~I.  Krastev, \emph{Phys.  Lett.  B}
\textbf{374} (1996) 1--6.  

\bibitem{fiorentini} 
G.~Fiorentini and B.~Ricci, \emph{Comments Mod.  Phys.  E} \textbf{1} (1999)
49--51.

\bibitem{bahcall2003} 
J.N.~Bahcall and C.~Pe{\~n}a-Garay, \emph{J.  High Energy Phys.} \textbf{11}
(2003) 004.

\bibitem{robertson} 
R.G.H.~Robertson, \emph{Prog.  Part.  Nucl.  Phys.} \textbf{57} (2006)
90--105.

\bibitem{waple} 
A.~Waple, \emph{Prog.  Phys.  Geog.} \textbf{23} (1999) 309.

\bibitem{crom} 
D.~Crommelynck, et~al., \emph{Geophys.  Res.  Lett.} \textbf{23} (1996)
2293.

\bibitem{spiro} 
See e.g.: M.~Spiro and D.~Vignaud, \emph{Phys.  Lett.  B} \textbf{242}
(1990) 279; N.~Hata, S.~Bludman and P.~Langacker, \emph{Phys.  Rev.  D}
\textbf{49} (1994) 3622; K.~M.  Heeger and R.~G.~H.  Robertson, \emph{Phys.  
Rev.  Lett.} \textbf{77}~(18) (1996) 3720--3723.

\bibitem{fogli2002} 
G.L.  Fogli, et~al., \emph{Phys.  Rev.  D} \textbf{66}~(5) (2002) 053010.

\bibitem{fogli2006} 
G.L.  Fogli, et~al., \emph{Prog.  Part.  Nucl.} \textbf{57} (2006) 742.

\bibitem{bahcall-pena} 
J.N.  Bahcall and C.~Pe{\~n}a-Garay, \emph{New Journal of Physics}
\textbf{6} (2004) 63; J.N.  Bahcall, \emph{Nucl.  Phys.  B (Proc.  Suppl.)} 
\textbf{118} (2003) 77.  

\bibitem{bahcall-gonzalez} 
J.~N.  Bahcall, M.~C.  Gonzalez-Garcia and C.~Pe{\~n}a-Garay, \emph{J.  High
Energy Phys.} \textbf{08} (2004) 016.

\bibitem{gonzalez} 
M.~Gonzalez-Garcia and M.~Maltoni, \emph{arXiv:0704.1800v1 [hep-ph]};
M.~Maltoni, T.~Schwetz, M.A.~Tortola and J.W.F.~Valle,
\emph{arXiv:hep-ph/0405172v6 Sept.  2007}.  

\bibitem{confortola} 
F.~Confortola et al., \emph {Phys.  Rev.  C} \textbf{75} (2007) 065803;
\emph{arXiv:0705.2151v1 [nucl-ex]} May (2007); T.A.  Brown et~al., \emph{arXiv:0710.1279v3 [nucl-ex]} Oct.  (2007).  

\bibitem{homestake98} 
B.T.~Cleveland, et~al.  (Homestake), \emph{Astrophys.  J.} \textbf{496}
(1998) 505--526; W.~Hampel, et~al.  (GALLEX), \emph{Phys.  Lett.  B}
\textbf{447} (1999) 127--133; J.N.~Abdurashitov, et~al.  (SAGE), 
\emph{Nucl.  Phys.  B} \textbf{118} (2003) 39; M.~G.  Altmann, et~al.  
(GNO), \emph{Phys.  Lett.  B} \textbf{616} (2005) 174; S.N.~Ahmed, et~al.
(SNO), \emph{Phys.  Rev.  Lett.} \textbf{92} (2004) 181301; Y.~Fukuda,
et~al.  (SuperK), \emph{Phys.  Lett.  B} \textbf{539} (2002) 179; T.~Araki,
et~al.  (KamLAND), \emph{Phys.  Rev.  Lett.} \textbf{94} (2005) 081801.

\bibitem{bahcall-2004-prl} 
J.N.~Bahcall and M.H.~Pinsonneault, \emph{Phys.  Rev.  Lett.}
\textbf{92}~(12) (2004) 121301.  

\bibitem{bakref} 
See Ref.~\cite{gonzalez} p.66--102 for a general discussion and extensive
references.

\bibitem{friedland} 
A.~Friedland, C.~Lunardini and C.~Pe\~na-Garay, \emph{Phys.  Lett.  B}
\textbf{594} (2004) 347; O.G.~Miranda, M.A.~Tortola and J.W.F.~Valle, 
\emph{J.  High Energy Phys.} \textbf{10} (2006) 008.  

\bibitem{fardon} 
R.~Fardon, A.E.~Nelson and N.~Weiner, \emph{J.  Cosmol.  Astropart.  Phys}
\textbf{0410} (2004) 005; M.~Cirelli, M.C.~Gonzalez-Garcia and 
C.~Pe\~na-Garay, \emph{Nucl.  Phys.  B} \textbf{719} (2005) 219; V.~Barger,
P.~ Huber and D.~Marfatia, \emph{Phys.  Rev.  Lett.} \textbf{95} (2005)
211802; M.C.  Gonzalez-Garcia, P.C.~Holanda and R.~Zukanovich Funchal,
\emph{Phys.  Rev.  D} \textbf{73} (2006) 033008.  

\bibitem{maltoni} 
M.~Maltoni, T.~Schwetz, M.A.~Tortola and J.W.F.~Valle, \emph{Phys.  Rev.  D}
\textbf{68} (2003) 113010.  

\bibitem{adams} 
J.S.~Adams, \emph{``Energy Deposition by Electrons in Superfluid Helium''},
Brown University PhD.  Dissertation (2001).

\bibitem{huang} 
Y.H.~Huang, \emph{``Solar Neutrino Detection Utilizing a Variant of Coded
Aperture on a Large Scale''}, Brown University PhD.  Dissertation (2007).


\bibitem{hillandketo} 
J.C.~Hill, O.~Heybey and G.K.~Walters, \emph{Phys.  Rev.  Lett.} \textbf{26}
(1971) 1213; W.~Stockton et al, \emph{Phys.  Rev.  Lett.} \textbf{24} (1970)
654.

\bibitem{maris} 
H.J.~Maris, \emph{J.  Low Temp.  Phys.} \textbf{87} (1992) 773; S.~Balibar,
\emph{Phys.  Lett.} \textbf{51A} (1975) 455; 
S.~Balibar et al, \emph{Phys.  Lett.} \textbf{60A} (1977) 135.

\bibitem{bandler} 
S.R.~Bandler et al, \emph{Phys.  Rev.  Lett.} \textbf{74} (1995) 3169;
S.R.~Bandler, \emph{``Detection of Charged Particles in Superfluid
Helium''}, PhD.  Dissertation, Brown University (1996); J.S.~Adams et al,
\emph{Phys.  Lett.  B} \textbf{341(3/4)} (1995) 431.  

\bibitem{fetter} 
A.L.~Fetter, in \emph{``The Physics of Liquid and Solid Helium'', Part I}.
John Willey and Sons, (1976) Sec.~2.3.4.

\bibitem{baskar} 
C.M.~Surko and F.~Reif, \emph{Phys.  Rev.} \textbf{175} (1968) 229;
B.~Sethumadhavan et al, \emph{Nucl.  Instr.  Meth.  A} \textbf{520} (2004)
142; B.~Sethumadhavan, \emph{``Charge Gain and Breakdown in Liquid Helium at
Low Temperatures''}, PhD.  Dissertation, Brown University (2007);
B.~Sethumadhavan et al, \emph{Phys.  Rev.  Lett.} \textbf{97} (2006) 015301.

\bibitem{fleischmann} 
A.~Fleischmann et al, \emph{Nucl.  Instr.  Meth.  A} \textbf{520} (2004) 27;
A.~Fleischmann et al, \emph{Nucl.  Instr.  Meth.  A} \textbf{444} (2000)
100; S.R.~Bandler et al, \emph{J.  Low Temp.  Phys.} \textbf{93} (1993) 709.

\bibitem{enss} 
A.  Fleischmann, C.  Enss and G.M.  Seidel, ``Metallic Magnetic
Calorimeters'' in \emph {Cryogenic Particle Detection}, Editor: C.  Enss,
Topics in Applied Physics Series, Springer Publ.  (2005).

\bibitem{bhaskar} 
B.  Sethumadhavan et al, \emph{Nucl.  Instr.  Meth.  A} \textbf{520} (2004)
142.

\bibitem{kim} 
Y.H.~Kim, \emph{``Thermodynamics of Low Temperature Detectors''}, Brown
University PhD.  Dissertation (2004); Y.H.~Kim et al, \emph{Nucl.  Instr.
Meth.  A} \textbf{520} (2004) 208.

\bibitem{seidel} 
G.M.  Seidel, R.E.  Lanou and W.  Yao, \emph{Nucl.  Instr.  Meth.  A}
\textbf{489} (2002) 189.  

\bibitem{reflect} 
\emph{Handbook of Optical Constants of Solids II}, Editor: E.D.  Palik, Academic Press (1991).  

\bibitem{geant} 
R.  Brun et al, \emph{Detector Description Program and Simulation Tool},
CERN, Geneva.  (1993).

\bibitem{lyons} 
L.~Lyons, \emph{Statistics for Nuclear and Particle Physicists}, Cambridge
University Press (1986); G.~Cowan, \emph{Statistical Data Analysis}, Oxford
University Press (1998).

\end{thebibliography}
\end{document}